\documentclass[aps,pre,twocolumn,superscriptaddress,showpacs]{revtex4}
\usepackage{dcolumn}
\usepackage{graphicx}
\usepackage[dvips]{hyperref}
\usepackage[TS1,OT1,T1]{fontenc}

\begin{document}

 \newtheorem{thm}{Theorm}
 \newtheorem{defn}{Definition}
 \newtheorem{cor}{collory}
 \newtheorem{lem}{lemma}
 \newtheorem{prop}{proposition}

% \preprint{}

\title{Cost-driven weighted networks evolution}



\author{Yihong Hu}
\email{051025007@fudan.edu.cn} \affiliation{Department of Management
Science, Fudan University, Shanghai 200433, China}

\author{Daoli Zhu}
\affiliation{Department of Management Science, Fudan University,
Shanghai 200433, China} \affiliation{The Institute of Shanghai
Logistics Development, Shanghai 200433, China}

\author{Yang Li}
\affiliation{The Institute of Shanghai Logistics Development,
Shanghai 200433, China}

\author{Bing Su}
\affiliation{Department of Management Science, Fudan University,
Shanghai 200433, China}

\author{Bingxin Zhu}
\affiliation{Department of Management Science, Fudan University,
Shanghai 200433, China}


\date{\today}

\begin{abstract}
Inspired by studies on airline networks we propose a general model
for weighted networks in which topological growth and weight
dynamics are both determined by cost adversarial mechanism. Since
transportation networks are designed and operated with objectives to
reduce cost, the theory of cost in micro-economics plays a critical
role in the evolution. We assume vertices and edges are given cost
functions according to economics of scale and diseconomics of scale
(congestion effect). With different cost functions the model
produces broad distribution of networks. The model reproduces key
properties of real airline networks: truncated degree distributions,
nonlinear strength degree correlations, hierarchy structures, and
particulary the disassortative and assortative behavior observed in
different airline networks. The result suggests that the interplay
between economics of scale and diseconomics of scale is a key
ingredient in order to understand the underlying driving factor of
the real-world weighted networks.
\end{abstract}

%\keywords{}

\maketitle

\section{Introduction}

Recently there is an increasing interest in studying weighted
networks dynamics. A wide array of models have been formulated to
capture various properties of real weighted networks
\cite{Boccalettia}. These models gradually take the social and
economic aspects into account in addition to physical properties. At
first some conditions in BA model \cite{BA1, BA2} are relaxed to
induce weight driven mechanism and weight dynamics coupling with
topology growth \cite{Antal, BBV1, BBV2, BBV3, DM, Wenxu Wang}. Then
for networks with spatial properties, physical distances are
included as weight and show its critical role in the evolution of
networks putting constraints on long-range links \cite{Kaiser,
GuimeraEPJB, Barratspatial, Yukio}. Recently noticing transportation
networks are designed with optimal objective, some models
\cite{Barratoptimal, Gastner1, Gastner2, Yanbo Xie} propose
different optimization policy either to minimize the whole system
cost or every traveler's cost and reproduce spatial structures in
real networks. Very recently human behavior on networks are
considered into the modeling \cite{yihong}. Most of these models
mention cost when explaining the rationality of the settings such as
disadvantaged long-rang links and optimal policies. However, they
don't explicitly model cost which in economics perspective is the
underlying driving factor of human behavior and the world.

Here we propose a model based on the theory of cost in
micro-economics which includes two rules: economy of scale and
diseconomy of scale. Economy of scale means reduction in cost per
unit resulting from increased production or service, while
diseconomy of scale is vice versa. Transportation networks are
designed and operated by companies with objectives to reduce cost
and improve profit. Therefore transportation networks obey the rules
in micro-economics. High traffic allows companies to use large
aircrafts and airports to utilize the facilities more frequently so
that the average operating cost is lowered. The hub-and-spoke
structure in airline networks also arise from economy of scale
because consolidation of passengers in different airplanes into one
aircraft at the hub can reduce operating cost greatly. Without this
economic benefit, the hub-and-spoke structures would not dominate
the airline industry for decades. On the other hand diseconomy of
scale also exist in transportation networks: high traffic over
capacity will bring extra cost due to congestion and increasing
management problems. The effect of congestion cost is evident in
transportation networks and communication networks as discussed in
\cite{Ashton, Jarrett, Danila1, Danila2, JJWu}. Unlike road networks
and communication networks, the congestion in airline networks only
occur at airports which have limited capacity. Busy flights make it
costly to obtain a new flight slot and easy to be delayed for
mistakes in scheduling and coordination.

With the objective of reducing cost and maximizing profit, airline
companies have to trade off between the economy of scale and
congestion effect when they want to establish new flights. The cost
reduction in connecting hub airports and high management fares for
congestion can be incorporated into one single cost function to
represent the average total cost. Our model assumes topological
growth and wight dynamics are both determined by cost adversarial
mechanism. With actual cost functions in airline industry our model
reproduces key properties of real airline networks: truncated degree
distribution, nonlinear strength degree correlation, hierarchy
structure, and particulary the disassortative and assortative
behavior observed in different airline networks. And comparing
different outcomes from different cost functions, we can draw the
conclusion that economy of scale introduces scale-free behavior and
diseconomy of scale is the reason for the cut-off in truncated
power-law distribution.

The rest of the paper is organized as follows: in section two, we
formulate the model and give concrete cost functions on vertices and
edges. In section three, we discuss degree evolution results
obtained by simulations and compare the simulation results from four
different cost functions. Section four analyzes the network
structure using cluster coefficient and degree correlations. The
last is conclusions.

%------------------------------------------------------------------------------
\section{The Model}
We assume the initial network has $N_0$ nodes with existing links.
There are initial weights $w_0$ on the links. The networks evolutes
according to the following rules:

\begin{itemize}
 \item {\it Growth}. At each step, a new node $n$ emerges with $m$ links.
 The probability of connecting old nodes $i$ is determined by:

\begin{equation}
\Pi_{n-i}=\frac{1/C_v(s_i)}{\sum_l 1/C_v(s_l)}
\end{equation}

where $C_v$, the cost of connecting vertex $i$, is a function of
strength $s_i$. This rule suggests that vertices with low cost have
large possibility to be selected.

\item {\it Weight dynamics}. All the connections are supposed to update their weights according to the following
probability:

\begin{equation}
w_{ij}\rightarrow\left\{
\begin{array}{lll}
w_{ij}+W, &\textrm{with probability} &p_{ij} \\
w_{ij}, &\textrm{with probability} &1-p_{ij}
\end{array}\right.
\end{equation}

where
\begin{equation}
p_{ij}=\frac{1/C_{e}(w_{ij})}{\sum_{s,t} 1/C_e(w_{st})}
\end{equation}

where $C_e$, the cost on edge, is a function of weight $w_{ij}$.
Edges with low cost have large probability to increase traffic.
\end{itemize}

There are two types of cost functions in the model: cost on vertices
and cost on edges.  Using different cost functions this model can be
used to simulate different networks such as airline networks and
supply chain networks. In airline networks cost on vertices is the
average total cost at airports including service fares and
congestion cost etc. Airline companies prefer airports with low cost
to open new flights. Cost on edges is the average operating cost of
aircrafts including fuel, pilots, maintenance cost etc
\cite{william}. The flight with low cost are more attractive to
passengers and airline companies will add more aircrafts for the
flight to meet the growing traffic. In supply chain networks the
cost on vertex represents the average production cost in firms and
the cost on connections represents average transaction cost between
firms. A firm would like to choose business partners with low
production cost and increase business interactions between firms
with low transaction cost.

Using appropriate cost functions, the model can recover some classic
models. Let $W=0$ and $C_v(s)=1/s$, the probability becomes
$\Pi_{n-i}=\frac{S_i}{\sum_l S_l}$. This is the classical strength
driven mechanism. And since the weight don't increase, the model
becomes equal to BA model. If $W=0$ and $C_v(s)$ is constant, the
old nodes are selected equally. The model produces random networks.
If $C_v(w_{ij})=1/s$ and $C_e(w_{ij})=s_is_j$, the model becomes the
traffic-driven model in \cite{Wenxu Wang}.

In the following we give cost functions in airline industry. The
cost function on vertices is:
\begin{equation}
C_{v}(s_{i})=s_i^{-1}+(s_i/s_0)^4
\end{equation}
where $s_0$ is the capacity of airports. The former part of the cost
function is decreasing with $s$ to represent economy of scale. The
latter part of cost function is increasing with $s$ representing
congestion cost according to well-known BPR (US Bureau of Public
Road) function and \cite{Philips}. The average operation cost at
airports first decrease with increasing traffic. When the traffic is
over the airport's capacity, congestion emerges and bring extra
cost. This function is analogous to the classical average production
cost function in micro-economics with the convex curve.

The cost function on edges is defined as:
\begin{equation}
C_{e}(w_{ij})=0.905w_{ij}^{-0.407}
\end{equation}
in which average aircraft operation cost is decreasing with traffic.
As we know congestion only occur at airports. So the cost on edge
only have economy of scale effect. According to \cite{Horner,
OKelly, william, Alan, Philips}, the total cost for operating a
flight is a Cobb-Douglas function $F=0.905w^{0.593}t^{0.744}$. $w$
is traffic volume and $t$ is stage length. Because we want to focus
on cost in this model, we ignore the physical distance $t$ and get
the above equation (5), leaving spatial properties to another paper,
. With the increase of traffic, average aircraft operation cost
decrease. This is so-called economy of scale in airline industry.

\section{Degree distribution}
\begin{figure}[th]
\begin{center}
\includegraphics[scale=0.8]{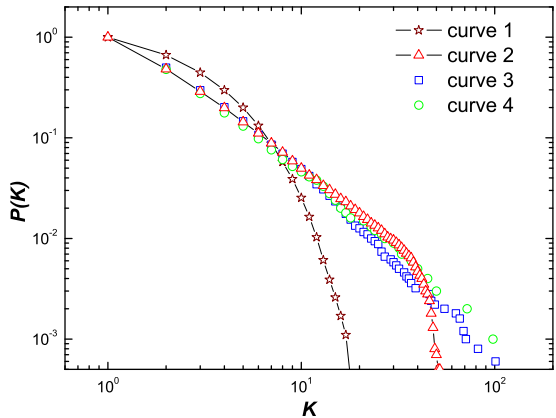}
\caption{Degree distribution. Curve 1: not power-law distribution
without economy of scale and congestion effect. Curve 2: truncated
power-law distribution with both economy of scale and congestion
effect. Curve 3: power-law distribution with economy of scale and no
congestion effect. Curve 4: power-law distribution with economy of
scale and no congestion. The data are averaged by 20 networks of
size $N=4000$ and $m=2$. \label{degreeplot}}
\end{center}
\end{figure}

The simulation result shows economy of scale contribute to
scale-free behavior and congestion effect introduces a cut-off into
scale-free degree distribution. See curve 2 in Fig.\ref{degreeplot}.
It's get under the model with the above mentioned cost functions (4)
and (5). The distribution decays as a truncated power-law:
\begin{equation}
P(K>k)\sim\left\{
\begin{array}{lll}
k^{-1.47}, &&k\leq30 \\
k^{-11.03}, &&k>30
\end{array}\right.
\end{equation}

All the airline networks reported in literature have such truncated
power-law degree distributions \cite{Wli, Bagler, GuimeraPNAS,
hongkun liu, Wangru, Dingding}. Several models are proposed to
explain it. The limit of connections to one airport is regarded as
the reason \cite{AmaralPNAS}. Spatial constraints are incorporated
to discourage long-rang links \cite{GuimeraEPJB, Barratspatial}.
However, spatial constraints are more appropriate to road networks
rather than to airline networks. Opening a direct flight is
convenient in airline networks since it does't require a physical
link. And if one flight can make money, even the longest links can
be set up as long as it's within the modern airline technology
capability. Geography distance is found insignificant when
topological ingredient plays a more important role \cite{Yanbo Xie}.
Therefore our model try to provide an explanation from inherent
economic perspective. To elaborate this point, we compare the
following four scenarios.

First, suppose there is no diseconomy of scale but only economy of
scale on edges and vertices. Let $C_v(s)=s^{-1}$,
$C(w_{ij})=0.905w_{ij}^{-0.407}$, $s_0=100$, and $W=1$. Run the
model and get the distribution as curve 4 in Fig. \ref{degreeplot}
which obey power-law distribution: $P(K>k)\sim k^{-r}$, $r=2.7$.

Second, we consider the case there are only economy of scale on
vertices with $C_v(s)=s^{-1}$, $s_0=100$, and $W=0$. In this
circumstance, the probability of connecting old vertices is
proportional to node strength and the weight is fixed. The degree
distribution is displayed as curve 3, a power-law with the exponent
a little small than curve 4.

\begin{figure}[th]
\begin{center}
\includegraphics[scale=0.8]{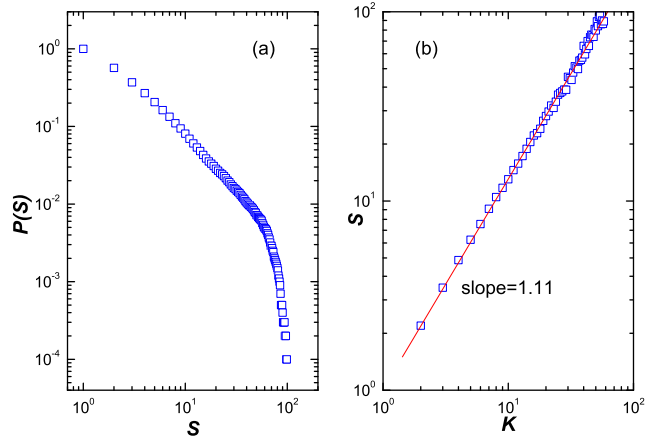}
\caption{Left: Strength distribution is a truncated power-law.
Right: strength $s_i$ vs. degree $k_i$. Linear data fitting gives
the slope 1.11, indicating the nonlinear correlation between
strength and degree $s \sim k^{1.11}$. The data are averaged by 20
networks of size $N=4000$, $m=2$ and using cost functions (4) and
(5) with $s_0=100$. \label{strength}}
\end{center}
\end{figure}

Comparing the curve 2, 3 and 4, we can see the congestion effect
contribute to the cut-off in the scale-free distribution and the
exponent of the power-law is influenced by the extent of economy of
scale. Double economy of scale on edges and vertices enhanced the
scale-free behavior in curve 3.

Finally, if there are either economy of scale or congestion effect,
i. e. $C_v(s)=$ constant, $C_v(s)=$ constant and $W=0$, the new
vertex has no preference in connecting old vertices and the weight
keeps fixed. We get random networks with the curve 1 obviously not a
power-law distribution. From the comparison between curve 1, 3 and
curve 4, we can see the scale-free behavior comes from economy of
scale

\section{Strength and degree}

Strength distribution and relation between strength and degree are
reported in Fig. \ref{strength}. The strength distribution also
display as a truncated power-law. And the average strength $s_i$ of
vertices with degree $k_i$ displays a non-trivial relations with
degree $k_i$ as confirmed by empirical observations in airline
networks of China, India, worldwide and North-America and other
transportation networks \cite{Andrea}. A large number of evolution
models of weighted networks produce linear relations between
strength and degree like the BBV model while only several models can
reproduce non-trivial relations \cite{Wenxu Wang, Yanbo Xie,
Barratspatial}. The nonlinearity in our model is induced by the
nonlinear cost change with the increasing traffic on edges and
vertices.

\section{Network structure}
\begin{figure}[th]
\begin{center}
\includegraphics[scale=0.8]{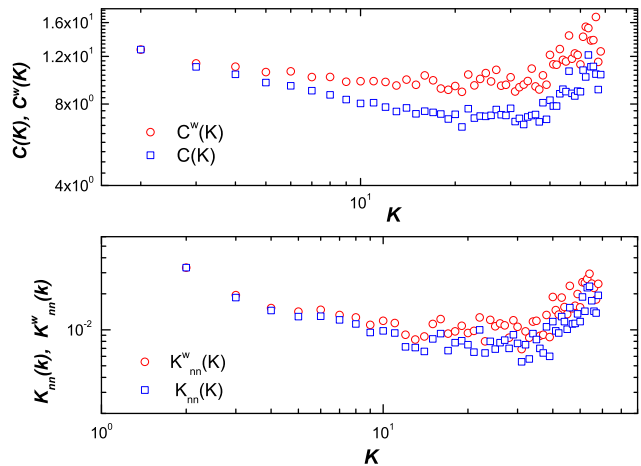}
\caption{Top: Unweighted and weighted clustering coefficient $C(k)$
and $C^w(k)$. Bottom: Unweighted and weighted average
nearest-neighbor degree $k_{nn}(k)$ and $k_{nn}^w(k)$. The data are
averaged by 20 networks of size $N=4000$ and $m=2$.
\label{clust_coeff}}
\end{center}
\end{figure}

To analyze the network structure, we calculate the cluster
coefficient and degree correlations in both pure topological and
weighted formulations defined in \cite{BarratPNAS}. The simulation
result is shown in Fig. \ref{clust_coeff}.

Both weighted cluster coefficient $c^{w}_i$ and weighted degree
correlations $k_{nn,i}^{w}$ are larger than cluster coefficient
$c_i$ and degree correlations $k_{nn,i}$ respectively, in good
agreement with the empirical data in airline networks. This result
reveals the existence of a rich-club phenomenon in that important
nodes form a group of well interconnected nodes. In other words
large vertices are well interconnected by large traffic edges. The
interconnected hubs have a major role in the network dynamics and
organization.

Degree correlations $k_{nn}(k)$ is a measure of assortativity. If
$k_{nn,i}$ is increasing with $k_i$, the network is called
assortative, i. e. large degree nodes tend to connect with large
degree nodes. While $k_{nn,i}$ is decreasing with $k_i$, the network
becomes disassortative.  Our simulation result $k_{nn}(k)$ and
$k_{nn}^w(k)$ decreases first in small degree then increases when
the degree becomes large. This means in small range the network
behaves disassortative but in large degree range the network behaves
assortative. This can be understood in the dynamic process in the
model. With small degree the diseconomy of scale doesn't show its
constraint on the connections to the hub. New node's connection to
hub is welcome and contributes to disassortativity. When the hub
becomes very large the congestion effect emerges and prevents
further connections from new nodes. But the traffic between hubs are
keeping growing because low cost from economy of scale attract
passengers and thus makes airline companies open more flights.

This result can explain some inconsistent results observed in real
airline networks. \cite{Bagler, hongkun liu, Dingding} shows
negative degree correlations in Chinese, India and Australian
airline networks, while \cite{Barratspatial} shows positive degree
correlations in Worldwide airline networks (WAN) and North American
networks. In our opinion this is due to the different scale of the
networks and the different extent of economy of scale and diseconomy
of scale. The airline networks of China, India and Australian are
relatively small and have less traffic compared with the worldwide
airline network and North-America network. Therefore they are more
influenced by economy of scale than diseconomy of scale. That's why
they display disassortative while WAN and North-America networks
show assortative.

\section{Conclusions}

In this paper we present a general model for weighted networks based
on the theory of cost in micro-economics including economy of scale
and diseconomy of scale. Our model reproduces key properties of
airline networks such as truncated power-law degree distribution,
non-linear relations between strength and degree, and hierarchy
structure. This model can produce broad distribution of networks if
we adopt appropriate cost functions. Particulary the disassortative
and assortative behavior observed in different airline networks can
be reproduced and explained by our model.

The main difference between our model and previous models is that
our model incorporates realistic economic factor into the modeling.
Airline networks are widely studied in economics, operations
research, engineering, transportation research and business
management. They are designed and operated to achieve optimal
objectives. Therefore airline networks evolution is quite different
from natural networks such as biological networks. Omit the
economics and human being behavior, we can't entirely understand the
airline networks evolution. Besides airline networks, the model can
also be used to simulate other networks with economic properties
such as supply chain networks. Our work discloses that the interplay
of economy of scale and congestion effect determines the evolution
of networks. We hope this model can enlighten our understanding
towards realistic weighted networks evolution.

%\appendix
%\section{}

%\begin{acknowledgments}
%\end{acknowledgments}

\begin{acknowledgments}

The work was supported by Natural Science Foundation of China (NSFC
70432001).

\end{acknowledgments}

\appendix

\end{document}